\documentstyle[twoside,fleqn,espcrc2,epsfig]{article}
 
\newcommand{\be}{\begin{equation}}
\newcommand{\ee}{\end{equation}}
\newcommand{\bea}{\begin{eqnarray}}
\newcommand{\eea}{\end{eqnarray}}

\newcommand{\nn}{\nonumber}

\title{Quenched Supersymmetry}
\author{A.~Donini\address{Dep. de Fisica Te\'orica, Universidad Aut\'onoma
de Madrid, Cantoblanco, E-28049, Madrid,Spain.} 
\thanks{Talk presented by A. Donini.}, 
E. Gabrielli$^{\rm a}$, M. B. Gavela$^{\rm a}$}

\begin{document}

\begin{abstract} 
 We study the effects of quenching in Super-Yang-Mills theory.
While supersymmetry is broken, the lagrangian acquires a new flavour
$U(1 \mid 1)$ symmetry. The anomaly structure thus differs from the
 unquenched case.
We derive the corresponding low-energy effective lagrangian. 
As a consequence, we predict the mass splitting expected in numerical
simulations for particles belonging to the lowest-lying supermultiplet.
\end{abstract}
 
\maketitle

\section{INTRODUCTION}

If it finally turns out that nature can express herself in the 
language of supersymmetric theories, the understanding of the strongly 
coupled regime of the latter may be very important. From gluino condensation
to many other issues in nowadays implementations of string theories,
such understanding is pertinent. A first important step in this direction
was made by Veneziano and Yankielowicz (VY) \cite{vy}, when they derived the
low energy effective lagrangian for pure $N=1$ supersymmetric Yang-Mills 
theory (SYM).  

The ``a priori'' primary tool for a direct study of strongly coupled field
theories is lattice regularization. Its numerical implementation can be
very time/resource consuming, and it is well known that the 
quenched approximation greatly reduces such costs, albeit at the price of 
the corresponding systematic error. Numerical results
for the spectrum of the lowest-lying supermultiplet
have been obtained in this approximation \cite{dghv}, 
and could be better justified by an analytical study of the effects 
of quenching in a SUSY theory. For recent unquenched results 
see \cite{mdg}.

In this talk we discuss the low energy effective lagrangian
for quenched $N=1$ supersymmetric Yang-Mills theory in the continuum \cite{dgg},
paralleling ref.~\cite{bg}.

\section{SYM AND ITS SYMMETRIES}

The SYM lagrangian is the simplest supersymmetric gauge theory
and describes a vector supermultiplet with fermion and boson fields
in the adjoint representation of the gauge group. No matter superfields
(containing fermion and scalar fields in the fundamental representation)
are present. The action is
\be
S_{SYM} = \int d x \left \{ - \frac{1}{4} F^a_{\mu\nu} F^{a\mu\nu}
 + \frac{i}{2} \bar \lambda^a \gamma^\mu D^{ab}_\mu \lambda^b \right \} ,
\ee
where $a= 1,2,3$ is the adjoint index and $D^{ab}$ is a covariant derivative.
This lagrangian, beyond gauge symmetry and supersymmetry, is classically 
invariant under chiral $U(1)$ transformation (a gluino mass term 
would also spoil supersymmetry). There is no vector $U(1)$ symmetry, 
since gluinos are Majorana fermions. Moreover, it obeys naive scale invariance.

However, chiral $U(1)$ and scale invariance are broken by the
chiral and trace anomalies, respectively:
\bea
\partial^\mu J_\mu &=& - c(g) F^a_{\mu\nu} 
\tilde F^{a\mu\nu} \, , \\
\Theta^\mu_\mu &=& c(g) F^a_{\mu\nu} F^{a\mu\nu} \,,
\eea
where $c(g) = \beta(g)/2 g$
and $\beta(g)$ is the $\beta$-function of the theory.
These two anomalies and the supersymmetric trace anomaly,
$\gamma^\mu S_\mu = 2 c(g) \sigma_{\mu\nu} F^a_{\mu\nu} \lambda^a$,
form the anomaly supermultiplet. 

Following ref.~\cite{dvv}, VY obtained the low energy effective action,
 $S_{VY}\, $\cite{vy}, by requiring that it reproduces the symmetries
 of the fundamental action, as well as its chiral, scale and 
supersymmetric anomalies:

\bea
S_{VY} &=& \int d^4 x \left \{ \frac{9}{\alpha} (S^\dagger S)^{1/3}_D 
\right . \\
& + & \left .
 \left [ \frac{1}{3} \left ( S \log \frac{S}{\mu^3} - S \right )_F + h.c. 
\right ] \right \} \nn ,
\eea
where $S$ is a chiral supermultiplet containing bound states of gluons
and gluinos (SYM is believed to be confined such as QCD), and $\alpha$
 and $\mu $ are free parameters.
The component fields of the supermultiplet $S$ aquire the common mass
$m_S = \frac{1}{3} \alpha \mu$ due to supersymmetry.

Recently, several modifications to this lagrangian has been proposed
\cite{ks}. We will study, however, the effect of quenching to
the VY lagrangian (modified theories could be studied in the same
approach). 

\section{qSYM AND ITS SYMMETRIES}

Paralleling ref.~\cite{bg},
we implement quenching at the fundamental level by adding to
the SYM lagrangian a ghost scalar field in the adjoint representation. 
The new action is:
\bea
S^q_{SYM} &=& \int d x \left \{ - \frac{1}{4} F^a_{\mu\nu} F^{a\mu\nu}
\right . \\
 &+& \left . \frac{i}{2} \bar \lambda^a \gamma^\mu D^{ab}_\mu \lambda^b 
 + \frac{i}{2} \bar \eta^a \gamma^\mu \gamma_5 D^{ab}_\mu \eta^b 
 \right \} \, . \nn
\eea
Unlike in \cite{bg}, the ghost kinetic operator is 
$\gamma_\mu \gamma_5 D^\mu$,
as a vector bilinear acting on commuting Majorana fields gives zero.
After the integration of the ghost degree of freedom in the functional
integral, the ghost and fermion determinants cancel each other,
 implementing the quenched approximation.
Introducing a set of generators $\sigma^i$ ($i = 0,\dots,3$)
mixing fermionic and bosonic fields, the action can be written as:
\be
S^q_{SYM} = \int d x \left \{ - \frac{1}{4} F^a_{\mu\nu} F^{a\mu\nu}
 + \frac{i}{2} \bar Q^a \sigma^0 \gamma^\mu_R D^{ab}_\mu Q^b 
 \right \}\, , 
\ee
where $Q$ is the doublet $Q^a=( \lambda^a , \eta^a )$. $\sigma^0$ is
the identity matrix, and $\sigma^{1,2,3}$ denote the Pauli matrices.
The action is no longer supersymmetric, as new bosonic fields
have been introduced with no fermionic counterparts. 
It is still gauge and classically scale invariant, though,
and its $U(1)$ chiral symmetry is promoted to a $Z_2$ graded
$U(1 \mid 1)$ chiral symmetry. The four currents associated to this symmetry
are:
\be
J^i_\mu = \bar Q^a \sigma^i \gamma^\mu_R Q^a \, .
\ee
Only the $J^3_\mu$ current is anomalous, while $J^0, J^1$ and $J^2$ are not:
 the gluino and ghost loops
give the same contribution to the anomaly (with opposite sign).
This means that the chiral anomaly breaks $U(1 \mid 1)$ to graded
 $SU(1 \mid 1)$.
The new anomalies of the theory are:
\bea
\partial^\mu J^3_\mu &=& - 2 c(g) F^a_{\mu\nu} 
\tilde F^{a\mu\nu}\, , \\
\Theta^\mu_\mu &=& \frac{\beta^\prime(g)}{\beta(g)} c(g) F^a_{\mu\nu} F^{a\mu\nu}\, ,
\eea
where $\beta^\prime$ is the $\beta$-function for the pure gauge theory
(the fermion and ghost contributions to the $\beta$-function cancel).

The low-energy effective lagrangian should be invariant under $SU(1 \mid 1)$
and reproduce these new anomalies. These type of requirement
fixed uniquely the supersymmetric low-energy theory, for the
lowest dimension operators, at leading order in the momentum expansion
\cite{vy}. Supersymmetry being absent in the present case, we briefly 
sketch below the method to derive the splitting in the mass
 spectrum induced by quenching.

Under the assumption that the $SU(1 \mid 1)$ chiral symmetry is
spontaneously broken by a singlet scalar field invariant under full
$U(1 \mid 1)$\footnote{This ansatz is suggested by a Coleman-Witten 
argument \cite{cw}, and supported by numerical results for quenched 
simulations \cite{dghv}.}, we consider as the new fields of the 
effective theory
\bea
\chi & = \sigma_{\mu\nu} F^a_{\mu\nu} Q^a & \chi \to U_R \chi \\ 
\phi^i & = \bar Q^a \sigma^i Q^a & \phi \to U_R \phi U_L^\dagger
\eea  
where $\phi = \sigma^i \phi^i = \frac{1}{\sqrt{2}} \rho \Sigma$, 
with $\rho$ a scalar field invariant under $U(1 \mid 1)$ and 
$\Sigma = e^{ i \hat \theta}$ a pseudoscalar field in the 
exponential representation ($\hat \theta = \theta^i \sigma^i, i = 0,\dots,3$).
$\theta_3$ is the $U(1 \mid 1)$ extension of the VY anomalous pseudoscalar,
whereas the three {\it pseudoscalar} fields $\theta^i$ ($i = 0,1,2$) 
are the Goldstone modes associated with the spontaneous symmetry 
breaking of $SU(1 \mid 1)$.

The low-energy effective lagrangian in terms of the component fields
is:
\bea
\label{qeff_resc}
{\cal L}_{qVY} &=& \frac{1}{2} \partial^\mu \sigma \partial_\mu \sigma 
+ \frac{1}{2} \left [ \partial^\mu \theta_0 \partial_\mu \theta_3 
+ \partial_\mu \theta^+ \partial^\mu \theta^- \right ] \nn \\
&+& \frac{1}{2} \partial^\mu \theta_3 \partial_\mu \theta_3 
+ \frac{i}{2} \bar \chi \gamma^\mu \partial_\mu \chi \nn \\ 
&-& \frac{c_0}{2} \mu \bar \chi \chi 
+ 4 \frac{\mu^2}{4 c_2} \theta_3^2 
- \frac{\mu^2}{4 d_2} \left ( \frac{\beta^\prime}{\beta} \right )^2 \sigma^2 
\nn \\
&+& {\rm interactions} + {\cal O} (1/ N_c^2)
\eea
where we write only the mass terms of the potential (since at this 
stage we are interested in the mass splitting). $c_0, \mu, c_2$ and $d_2$
 are free parameters. To obtain this
lagrangian several steps were performed:
\begin{itemize}
\item
 To impose that the most general lagrangian invariant under $U(1 \mid 1)$
is naively scale invariant, paralleling the analysis for QCD \cite{gjjs},
 and that the
only $U(1 \mid 1)$ and scale breaking terms are logarithms,
whose transformations give rise to the right anomalies, alike to \cite{dvv}.
\item
Only terms up to $(F^a_{\mu\nu}F^{a\mu\nu})^2$, 
$(F^a_{\mu\nu} \tilde F^{a\mu\nu})^2$ in 
the scalar potential are retained, based in $1/N_c$ 
arguments\footnote{ As usual, $F^a_{\mu\nu}F^{a\mu\nu}$ is an auxiliary
 field in the lowest order Lagrangian\cite{dvv}.} .
\item
The algebraic equations of motion for the 
auxiliary fields $F^a_{\mu\nu}F^{a\mu\nu}, 
F^a_{\mu\nu} \tilde F^{a\mu\nu}$ were used, in order to express
 the on-shell lagrangian
in terms of the physical fields $\chi, \phi$ only.
\item 
To expand around true minimum, so as to obtain the mass term of the potential.
The relevant fields are the scalar $\sigma$, the {\it pseudoscalars} 
$\theta_0, \theta_3$ and $\theta^\pm$ and the {\it fermions}
$\chi_\lambda,\chi_\eta$. Recall that $\theta^\pm$ and $\chi_\eta$ 
obey fermionic and bosonic statistics, respectively.
\end{itemize}

The quadratic term and the kinetic term for the $\theta_3$ field
are interpreted as vertices, since in the quenched theory
no resummation of the mass terms can be done \cite{bg}.

We can perform the limit $\eta \to 0$ in order to
recover the VY theory. 
In this limit,
\be
c_0 = \frac{1}{4 c_2} = \frac{1}{4 d_2} = \frac{1}{3} \alpha \mu\, .
\ee
When performing this limit the coefficients of the operators
were assumed to be analytical in the ghost field dependence, 
but for the coefficients of the anomalous terms, which are not. 
This non-analiticity is responsible for the mass splitting of the 
VY supermultiplet.

The mass spectrum for the bound states of gluino and gluon
fields is given by:
\bea
m_\sigma &=& \frac{\beta^\prime}{\beta} m_\chi \, ,\nn \\
m_\chi   &=& \frac{1}{3} \alpha \mu\,, \\
m_\theta &=&  2 m_\chi \,,\nn
\eea
to be compared with $m_\chi \,=\, m_\sigma \,=\, m_\theta$ in the unquenched
theory. The numerical results of \cite{dghv} are in fair agreement
with the predicted $\sigma-\chi$ mass splitting for SU(2), 
although an accurate measure of the quenched non-OZI contribution
to the $\theta$ mass is still needed.


\section*{Acknowledgements.}

We thank A. Gonzalez-Arroyo, P. Hernandez, G. C. Rossi, M. Testa
and A. Vladikas for useful discussions.


\begin{thebibliography}{10}
\bibitem{vy}
G. Veneziano and S. Yankielowicz, Phys. Lett. {\bf B113} (1982) 231.
\bibitem{dghv}
G.~Koutsoumbas and I.~Montvay, Phys. Lett. {\bf B398} (1997) 130;
A. Donini {\em et al.}, Nucl. Phys. {\bf B523} (1998) 529.
\bibitem{mdg}
DESY-M\"unster, hep-lat/9808024.
\bibitem{dgg}
A. Donini, E. Gabrielli and M. B. Gavela, in preparation.
\bibitem{bg}
C. Bernard and M. Golterman, Phys. Rev. {\bf D46} (1992) 853.
\bibitem{dvv}
P. Di Vecchia and G. Veneziano, Nucl. Phys. {\bf B171} (1980) 253.
\bibitem{ks}
A. Kovner and M. Shifman, Phys. Rev. {\bf D56} (1997) 2396;
G. Farrar {\em et al.}, Phys. rev. {\bf D58} (1998) 015009;
hep-th/9806204.
\bibitem{cw}
S. Coleman and E. Witten, Phys. Rev. Lett. {\bf 45} (1980) 100.
\bibitem{gjjs}
H. Gomm {\em et al.}, Phys. Rev. {\bf D33} (1986) 801.
\end{thebibliography}
\end{document}